\documentclass[11pt,twosize]{article}
\usepackage{asp2010}

\resetcounters

\begin{document}

\title{Spotting Radio Transients with the help of GPUs}
\author{Benjamin~R~Barsdell$^1$, Matthew~Bailes$^1$, David~G~Barnes$^2$ and
  Christopher~J~Fluke$^1$ 
  \affil{$^1$Swinburne University of Technology \\ PO Box 218
    \\ Hawthorn VIC 3122 (Mail H39) \\ Australia \\
         $^2$ Monash e-Research Centre \& Life Sciences Computation Centre \\
	Bldg 220, 770 Blackburn Rd \\ Monash University VIC 3800 \\ Australia}
}

\begin{abstract}
Exploration of the time-domain radio sky has huge potential for advancing our
knowledge of the dynamic universe. Past surveys have discovered large numbers
of pulsars, rotating radio transients and other transient radio phenomena;
however, they have typically relied upon off-line processing to cope with the
high data and processing rate. This paradigm rules out the possibility of
obtaining high-resolution base-band dumps of significant events or of
performing immediate follow-up observations, limiting analysis power to what
can be gleaned from detection data alone. To overcome this limitation,
real-time processing and detection of transient radio events is required. By
exploiting the significant computing power of modern graphics processing units
(GPUs), we are developing a transient-detection pipeline that runs in
real-time on data from the Parkes radio telescope. In this paper we discuss
the algorithms used in our pipeline, the details of their implementation on
the GPU and the challenges posed by the presence of radio frequency
interference.
\end{abstract}

\section{Introduction}
The High Time Resolution Universe (HTRU) survey currently underway at Parkes
Observatory using the 64m dish is an all-sky search for transient and periodic
point sources. Its primary goals are to discover new pulsars and fast radio
transients and it has so far discovered more than 65 previously unknown
pulsars since it began in 2008 \citep{KeithEtal2011}. The observing backend
has 400 MHz of bandwidth over 1024 channels centered at 1381.8 MHz, and data
products are produced with a time resolution of 64$\mu s$. The receiver is a
multibeam design and contains 13 separate feed horns pointing at nearby
locations on the sky \citep{StaveleySmith1996}.

The current pulsar and transient search process begins by taking filterbank
data (i.e., time series for each frequency channel) from the backend -- stored
using 2 bits per sample to give a data rate of around 4 MB/s per beam -- and
sending it via a fibre optic link from the telescope down to the Swinburne
supercomputer in Melbourne ($\sim$ 700 km away). There it is written to disk
for immediate processing (when possible) and also to tapes for longer term
storage and future (re-)processing. The processing itself is then performed on
a distributed cluster of multi-core central processing units (CPUs). Once the
data are in memory, the current transient search pipeline takes more than 30
minutes to process a 10 minute observation.

While limitations in CPU computing power have necessitated offline processing
of the survey data to date, ideally one would like to execute the pipeline in
\textit{real-time} with the observations. If this were achieved, several
significant advances become possible. Among them is instant feedback and
capacity for follow-up observations of significant events. This would
eliminate the long delays incurred by the current off-site processing and
undoubtedly increase the survey's discovery power. Another promising
possibility is triggered baseband data dumps. These would allow the capture of
full-resolution baseband data during events of interest; the baseband data are
currently not saved due to the high data rate, but could be kept for a short
period in a buffer that is written to disk when deemed worthwhile.

CPU computing performance continues to increase year on year, but is still not
up to the task of processing the survey data in real-time (for acceptable
monetary, power and floorspace costs). However, the recent appearance of
graphics processing units (GPUs) in high-performance computing presents a new
opportunity to attempt to break the real-time barrier. These devices can
provide an order of magnitude more compute power than CPUs at comparable
costs, but pose considerable software challenges due to their unfamiliar
architectures [see \citet{FlukeEtal2011} for an introduction]. It is our aim
to implement the transient detection pipeline on a GPU and to exploit the
boost in processing power to achieve a real-time processing rate.

\section{The radio-transient detection pipeline}
The transient detection part of the software pipeline, which begins with
filterbank data and ultimately produces a list of candidate events, involves
five processes: RFI mitigation, incoherent dedispersion, baseline/red-noise
removal, matched filtering and peak finding. With a special focus on the first
two, we now introduce these algorithms and describe their implementation on
the GPU. We base our approach on the algorithm analysis methodology of
\citet{BarsdellEtal2010}.

\subsection{RFI mitigation}
When searching for transient events in real time, one of the most significant
obstacles to overcome is the presence of radio frequency interference
(RFI). An event detection and recording system could easily be swamped with
false positives if significant levels of RFI are allowed to pass through the
pipeline. It is therefore necessary to apply RFI mitigation techniques to
reduce the effects of these undesirable phenomena.

Several RFI rejection techniques have been used in the literature. These
include simple sigma-clipping (where unnaturally loud signals are excised),
statistical methods such as the use of spectral kurtosis (where
non-Gaussianity is detected and removed) \citep{NitaEtal2007} and coincidence
rejection via the use of one or more reference antennas (where non-localised
signals are assumed to be RFI) \citep{FridmanBaan2001}. In this work we
implement the coincidence rejection method using the 13 beams of the Parkes
multibeam receiver as an array of reference antennas.

The simplest approach to implementing the coincidence rejection technique is
to apply signal-to-noise and coincidence thresholds such that a) there is a
good distinction between true signals and RFI, and b) there is a satisfactory
constraint on the probability of falsely classifying noise spikes as RFI. An
example of such thresholds might be ``a signal exceeding 3$\sigma$ in 4 or
more beams'', which gives a probability of falsely classifying a noise sample
as RFI of $p \le {13 \choose 4} \times 0.01^4 = 7.15 \times 10^{-6}$. This is
the method we chose for the initial version of the pipeline. The algorithm
takes the form of a \textit{transform} [see \citet{BarsdellEtal2010}] of the
multibeam time series, where for each sample the 13 beams are checked against
the thresholds and the result is written as a boolean RFI mask. This maps
trivially to the GPU as an ``embarassingly parallel'' problem.

\subsection{Incoherent dedispersion}
For transient search pipelines, dedispersion is typically the most
computationally intensive process. The algorithm arises from the need to
counteract the effect of the frequency-dependent time delay induced into the
signal by interactions with the interstellar medium. This delay (or
\textit{smearing}) increases with the number of free electrons between us and
the source. Given that the distance to new sources remains unknown, the amount
of dispersion (the \textit{dispersion measure}, DM) in the signal must be
guessed prior to executing the remainder of the pipeline. Surveys typically
compute many trial DMs -- approximately 1200 in the case of the HTRU
survey. The computation of each dispersion trial involves a complete
integration over frequency channels, so it is clear that the process in its
entirety requires a significant amount of computation.

While computationally intensive, the incoherent dedispersion algorithm is
relatively simple to implement. Furthermore, as identified by
\citet{BarsdellEtal2010}, the properties of the algorithm make it a very good
match for the architecture of a GPU. Indeed, efficient implementations using
NVIDIA's CUDA\footnote{http://www.nvidia.com/object/cuda\_home\_new.html} GPU
programming platform have appeared in the literature with reported speed-ups
of an order of magnitude or more over multi-core CPU codes
[\citet{MagroEtal2011}, \citet{O02_adassxxi}, Barsdell et al. (MNRAS
submitted)].

Our GPU implementation of the most common `direct' dedispersion algorithm [see
Barsdell et al. (refereed) for details] has reduced the processing time
from around 20 minutes to 2.5 minutes for a 10 minute observation. While a
significant speed-up in its own right, the more important aspect of this
result is the fact that it has broken through the real-time barrier. Given
that dedispersion consumes the largest fraction of the total execution time,
the sub-real-time GPU implementation is a critical element for the real-time
pipeline.

\subsection{Other algorithms}
The recorded time series from the telescope often contain baseline wiggles
(red noise), and these must be removed to ensure consistent behaviour in the
detection pipeline. In the time domain, the most common approach is to compute
a running mean for the data with a given window size (equivalent to convolving
with a wide boxcar) and then subtracting this off the original data. One way
to implement a running mean efficiently on a GPU is to exploit a parallel
prefix sum algorithm. Once the `running sum' has been computed, the running
mean may be calculated by subtracting the points at plus and minus the window
radius. We have implemented this algorithm using the Thrust library, which
provides optimised implementations of the prefix sum and transform algorithms
(among many others).\footnote{http://code.google.com/p/thrust/}

In order to detect signals with a range of widths, the pipeline performs a
series of matched filtering operations. This involves convolving the time
series with boxcar functions of increasing sizes (e.g., 2, 4, 8
etc. samples) and passing each result through the remainder of the
pipeline. The matched filtering algorithm is essentially identical to the
baseline removal algorithm, and can be implemented in the same manner using
the parallel prefix sum, which we have again done using the Thrust library.

The final step of the pipeline is to detect peaks in the reduced time
series. We have implemented this process by first performing a sigma-cut and
then gathering contiguous regions together into individual events. These
algorithms map to the `transform' and `reduce\_by\_key' functions of the
Thrust library, which makes their implementation on the GPU a trivial matter.

\section{Discussion}
Having successfully implemented each step of the transient detection process
on a GPU, we are now working to integrate them into a complete GPU
pipeline. Early benchmarks indicate that our target of processing one beam per 
GPU in real time is well within reach. Once the software pipeline is
completely operational, we plan to deploy it on a small cluster of GPU nodes
on-site at the Parkes radio telescope.

Our real-time radio transient detection pipeline promises to simplify the data
processing procedure by performing it as observations are made. This new data
reduction paradigm will enable immediate follow-up observations upon the
detection of interesting transient events and provide unprecedented resolution
of unique phenomena via baseband dumps.


\bibliography{abbrevs,O05}

\end{document}